\documentclass[aps,superscriptaddress,twocolumn,twoside,floatfix,pra,nofootinbib,a4paper]{revtex4-2}

\pdfoutput=1

\usepackage{times}
\usepackage{dsfont}
\usepackage{bm} 
\usepackage{amsfonts}
\usepackage{amsmath}
\usepackage{amssymb}
\usepackage{amsthm}
\usepackage{color}
\usepackage{enumerate}
\usepackage{multirow}
\newcommand{\stkout}[1]{\ifmmode\text{\sout{\ensuremath{#1}}}\else\sout{#1}\fi}
\usepackage{latexsym}
\usepackage{mathrsfs}
\usepackage{verbatim}
\usepackage{float}
\usepackage{pifont}
\usepackage{caption}
\usepackage{empheq}
\usepackage{IEEEtrantools}

\usepackage{microtype}

\usepackage[dvipsnames]{xcolor}

\usepackage{natbib}
\usepackage[colorlinks=true,linkcolor=blue,citecolor=magenta,urlcolor=blue]{hyperref}

\usepackage{cleveref} 

\usepackage[many]{tcolorbox}

\newtcolorbox[auto counter]{framefloat}[2][]{title=Box~\thetcbcounter: #2,,fonttitle=\bfseries, boxsep=0mm,boxrule=1pt,colframe=black,colback=white,coltitle=black,float=t!,#1}

\definecolor{maroon}{cmyk}{0,0.87,0.68,0.32}

\DeclareMathOperator{\Tr}{tr}

\usepackage{physics} 

\newcommand{\expect}[1]{\langle#1\rangle}

\usepackage{array,mathtools,amssymb,booktabs}
\newcolumntype{C}{>{$}c<{$}}
\AtBeginDocument{
\heavyrulewidth=.08em
\lightrulewidth=.05em
\cmidrulewidth=.03em
\belowrulesep=.65ex
\belowbottomsep=0pt
\aboverulesep=.4ex
\abovetopsep=0pt
\cmidrulesep=\doublerulesep
\cmidrulekern=.5em
\defaultaddspace=.5em
}

\usepackage{epsfig}
\usepackage{graphicx}
\graphicspath{{figures/}}
\usepackage{subfig}
\usepackage{caption}
\begin{document}

\title{Projective measurements are sufficient for recycling nonlocality}

\author{Anna Steffinlongo}
\affiliation{Dipartimento di Fisica e Astronomia ``G.Galilei",	Universit\`a degli Studi di Padova, I-35131 Padua, Italy}
\affiliation{Institute for Quantum Optics and Quantum Information -- IQOQI Vienna, Austrian Academy of Sciences, Boltzmanngasse 3, 1090 Vienna, Austria}
\affiliation{Atominstitut,  Technische  Universit{\"a}t  Wien, Stadionallee 2, 1020  Vienna,  Austria}

\author{Armin Tavakoli}
\affiliation{Institute for Quantum Optics and Quantum Information -- IQOQI Vienna, Austrian Academy of Sciences, Boltzmanngasse 3, 1090 Vienna, Austria}
\affiliation{Atominstitut,  Technische  Universit{\"a}t  Wien, Stadionallee 2, 1020  Vienna,  Austria}

\date{\today}

\begin{abstract}
Unsharp measurements are widely seen as the key resource for recycling the nonlocality of an entangled state shared between several sequential observers. Contrasting this, we here show that nonlocality can be recycled using only standard, projective, qubit measurements. Focusing on the CHSH inequality and allowing parties to share classical randomness, we determine the optimal trade-off in the magnitude of Bell violations for a maximally entangled  state. We then find that  non-maximally entangled states make possible larger sequential violations, which contrasts  the standard CHSH scenario. Furthermore, we show that nonlocality can be recycled using projective qubit measurements even when no shared classical randomness is available. We discuss the implications of our results for experimental implementations of sequential nonlocality.
\end{abstract}

\maketitle

\textit{Introduction.---} In the simplest, Clauser-Horne-Shimony-Holt (CHSH) \cite{CHSH1969}, Bell experiment, two parties perform local measurements on a shared pair of entangled qubits. Regardless of which pure state they share, the largest violation of the CHSH inequality is  obtained using standard measurements, corresponding to basis projections. Such measurements maximally perturb the state, rendering it separable after the measurement. Nevertheless, this is not a concern in standard Bell scenarios because each qubit is  measured only once.

\begin{figure}
	\centering
	\includegraphics[width=\columnwidth]{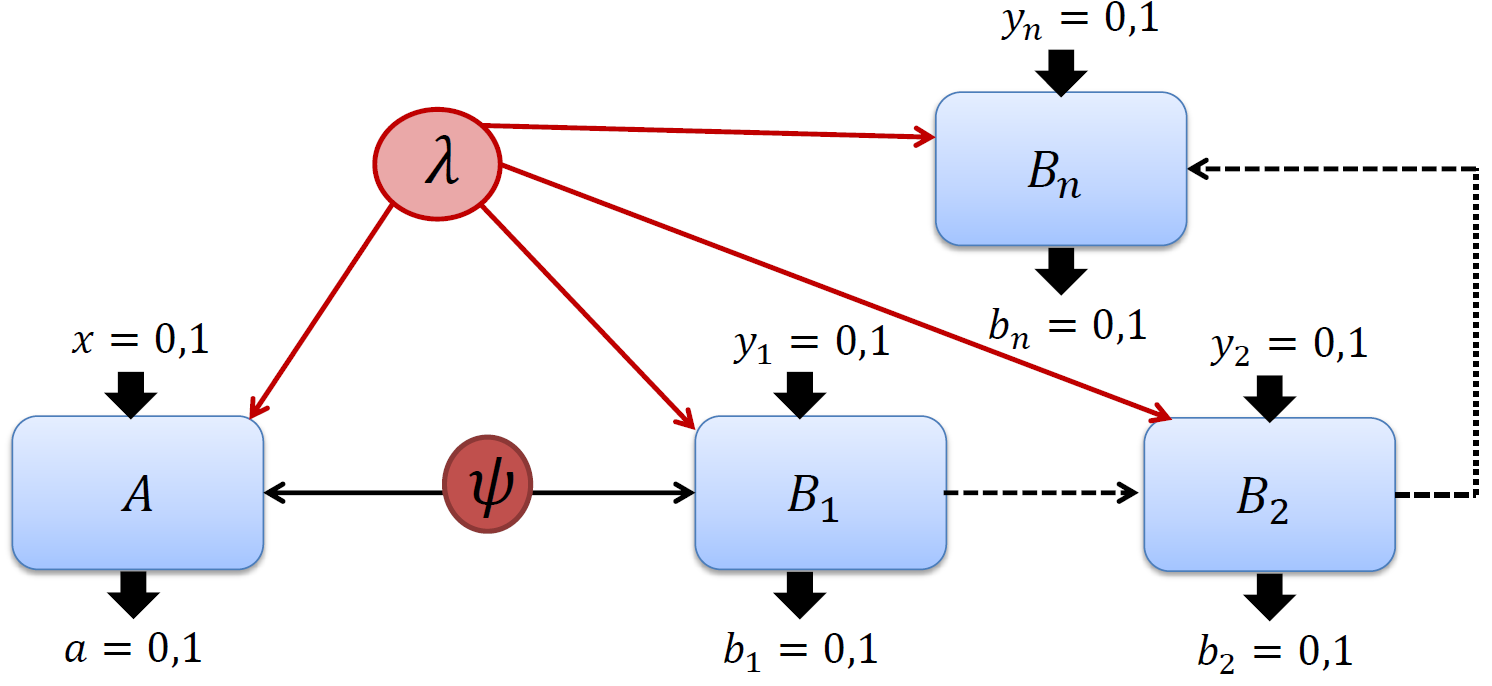}
	\caption{A source distributes a two-qubit state $\psi$ to measuring parties $A$ and $B_1$. The qubit of $B_1$ is sequentially relayed, measured, relayed etc, until measured by a final party $B_n$. Before the experiment begins, all parties may agree on sharing strings of classically correlated data $\lambda$.}\label{FigScenario}
\end{figure}

Beginning with the work of Silva \textit{et.~al.,~}\cite{Silva2015}, there has in recent years been much interest in whether the post-measurement state in a Bell experiment can be re-used for showcasing nonlocality between several observers who perform sequential measurements. This has been explored in both theory \cite{Mal2016, Cabello2018, Kumari2019, Saha2019, Das2019, Bowles2020, Curchod2017, Brown2020, Cabello2021, Cheng2021, Chengb2021, Zhang2021} and experiment \cite{Schiavon2017, Hu2018, Calderaro2020, Feng2020, Foletto2021}, and it has inspired similar investigations of other quantum correlation tasks (see e.g.~\cite{Bera2018, Sasmal2018, Choi2020, Anwer2021, Mohan2019, Miklin2020, Foletto2020, Anwer2020}). In such scenarios, a two-qubit state is shared between measuring parties $A$ and $B_1$. The post-measurement state of $B_1$ is relayed to another, independent, measuring party $B_2$. The relay process is continued until the qubit is measured by the final sequential party $B_n$ (see Figure~\ref{FigScenario}). The aim is for every pair $A-B_k$ (for $k=1,\ldots,n$) to violate a Bell inequality. To this end, each sequential party (except $B_n$) must perform a compromise measurement: it must produce strong enough correlations with $A$ to elude local models, but still preserve enough of the entanglement between the qubits to enable the next party to do the same. 

In recycling protocols, the standard procedure is to let the measurement apparatus interact weakly with the incoming qubit. It corresponds to an unsharp measurement, which can be realised by  L\"uders type quantum instruments \cite{Pellon2012}. By tuning $B_k$'s sharpness, one obtains a trade-off between the magntiude of the Bell parameter observed between $A-B_k$ and the amount of nonlocality left for the remaining pairs $A-B_{j}$, for $j>k$, to consume. %

Here, we depart from unsharp measurements and consider the exclusive use of standard projective measurements for recycling violations of the CHSH inequality between independent parties measuring one share of a two-qubit state. Provided the intuition that useful projective measurements disentangle a state, which commonly motivates the use of unsharp measurements, our endeavour may at first appear futile. Indeed, a CHSH test, in which each party has two measurements, offers only three different classes of projective qubit measurement strategies. (i) Both measurements are basis projections. This renders the state separable, preventing a second violation. (ii) Both  measurements are trivial, each corresponding to identity measurements, preventing a first violation. (iii) One measurement is a basis projection and the other is trivial: since one input is effectively discarded, it prevents a first violation.

Our approach is to overcome the apparent inabilities of projective measurement strategies by leveraging classical randomness. We use shared randomness to  stochastically combine different, individually unsuccessful, projective measurement strategies to achieve three sequential violations of the CHSH inequality. For two sequential violations, we analytically characterise the optimal trade-off between the two CHSH parameters, as obtained from  a maximally entangled state. We then consider pure partially entangled two-qubit states and show that these make possible larger double violations of the CHSH inequality. This stands in contrast to quantum nonlocality in the standard CHSH scenario\cite{Horodecki1995}, where such states are strictly weaker than the maximally entangled state. We then proceed to investigate whether shared randomness is necessary in order to facilitate sequential violations under projective measurements. Interestingly we answer this in the negative, which leads us to protocols that are arguably minimally resource-consuming for recycling nonlocality. We discuss the outlook of our results in the context of experimental implementations and highlight some open problems on the use of projective measurements to recycle quantum nonlocality.

\textit{Sequential scenario.---} While we will mainly focus on the simplest sequential scenario, we introduce it in its general form (see Figure~\ref{FigScenario}). A two-qubit state $\psi$ is shared between parties $A$ and $B_1$. They each privately select one of two inputs, denoted $x\in\{0,1\}$ and $y_1\in\{0,1\}$ respectively, and perform a corresponding quantum measurement. The outcomes are denoted $a\in\{0,1\}$ and $b_1\in\{0,1\}$. Party $B_1$ then relays the post-measurement qubit to party $B_2$ who similarly selects $y_2\in\{0,1\}$, performs an associated measurement, outputs $b_2\in\{0,1\}$ and relays the post-measurement qubit. This process continues until the final sequential party, $B_n$, selects $y_n\in\{0,1\}$ and outputs $b_n\in\{0,1\}$. Before the experiment begins, the parties may also agree to share correlated strings of classical data, $\lambda$, subject to some probability distribution $\{p(\lambda)\}_\lambda$. All parties select their inputs without bias and the sequential parties act independently: $B_k$ relays no classical information about $(y_k,b_k)$ to $B_{k+1}$.

Each pair $A-B_k$ tests the CHSH inequality, 
\begin{equation}\label{CHSH}
S_k\equiv\sum_\lambda p(\lambda) S_k^{(\lambda)} \leq 2,
\end{equation}
where  $S_k^{(\lambda)}\equiv  \sum_{x,y_k}(-1)^{xy_k} \expect{A^{(\lambda)}_x,B^{(\lambda),k}_{y_k}}_{\psi^{(\lambda)}_k}$. Here,  $\{A^{(\lambda)}_x,B^{(\lambda),1}_{y_1},\ldots,B^{(\lambda),n}_{y_n}\}$ denote the observables of the respective parties conditioned on $\lambda$ and the expectation value is evaluated with respect to the average recycled state $\psi^{(\lambda)}_k$ (with $\psi^{(\lambda)}_1=\psi$). This state is recursively given by
\begin{equation}\label{poststate}
\psi^{(\lambda)}_{k+1}=\frac{1}{2}\sum_{y_k,b_k}\left(\openone\otimes  K^{(\lambda),k}_{b_k|y_k}\right) \psi^{(\lambda)}_{k}\left(\openone\otimes  K^{(\lambda),k}_{b_k|y_k}\right)^\dagger,
\end{equation}
where $\{K^{(\lambda),k}_{b_k|y_k}\}$ are the Kraus operators representing the quantum instrument used by $B_k$ to realise the measurement  $B^{(\lambda),k}_{b_k|y_k}=\left(K^{(\lambda),k}_{b_k|y_k}\right)^\dagger K^{(\lambda),k}_{b_k|y_k}$ when advised by $\lambda$.

We are interested in quantum protocols based on projective qubit measurements.  Thus, $B^{(\lambda),k}_{b_k|y_k}=\left(B^{(\lambda),k}_{b_k|y_k}\right)^2$. The Kraus operators then take the form $K^{(\lambda),k}_{b_k|y_k}=U_{b_ky_k}^{(\lambda),k}B^{(\lambda),k}_{b_k|y_k}$, where $U_{b_ky_k}^{(\lambda),k}$ are arbitrary unitary operators. This means that a party first measures projectively and then performs a unitary based on the input and output. Notice that for  qubits, all projective measurements are either basis measurements, i.e.~they correspond to measuring in the direction of some Bloch vector ($\{\ket{\vec{n}},\ket{-\vec{n}}\}$), or trivial identity projections, for which the outcome is independent of the state ($\{\openone,0\}$).  When performed on one share of an entangled qubit pair, the former renders the post-measurement state separable while the latter leaves it unchanged.

For the simplest scenario, namely $n=2$, which is our main focus, we simplify the notations by calling the parties $\{A,B,C\}$, their inputs $\{x,y,z\}$, their outcomes $\{a,b,c\}$, their observables $\{A^{(\lambda)}_x,B^{(\lambda)}_{y},C^{(\lambda)}_{z}\}$ and $B$'s unitaries $U_{by}^{(\lambda)}$.

\textit{Optimal trade-off for maximally entangled state.---} Consider the case of $n=2$ when the shared state is maximally entangled, $\ket{\psi}=\ket{\phi^+}=\frac{1}{\sqrt{2}}\left[\ket{00}+\ket{11}\right]$.  We set out to determine the set of possible pairs of CHSH parameters, $(S_1,S_2)$, reachable under projective measurements and shared randomness. To this end, we must determine the optimal trade-off, i.e.~the largest possible value of $S_2$ given a value of $S_1$. 

Since $A$'s qubit is only measured once, the optimal measurements are rank-1 projective. These can w.l.g.~be written as the observables $A_x^{(\lambda)}=\cos\theta^{(\lambda)}\sigma_X+(-1)^x\sin\theta^{(\lambda)}\sigma_Z$.
Then, using that  $O\otimes\openone \ket{\phi^+}=\openone\otimes O^\text{T}\ket{\phi^+}$ for any linear operator $O$, we can for given $\lambda$ write the first CHSH parameter as 
\begin{equation}\label{S1}
S_1^{(\lambda)}=\cos\theta^{(\lambda)}\Tr\left(\sigma_X B^{(\lambda)}_0 \right)+\sin\theta^{(\lambda)}\Tr\left(\sigma_Z B^{(\lambda)}_1 \right).
\end{equation}
Using the state update rule \eqref{poststate} for projective measurements, and the completeness condition $C_{1|z}=\openone-C_{0|z}$, we write the second CHSH parameter as
\begin{align}\nonumber\label{S2}
S_2^{(\lambda)}&=\cos\theta^{(\lambda)} \sum_{y,b} \Tr\left(B_{b|y}^{(\lambda)}\sigma_X B_{b|y}^{(\lambda)} \left(U^{(\lambda)}_{by}\right)^\dagger C_{0|0}^{(\lambda)}U_{by}^{(\lambda)}\right)\\
&+\sin\theta^{(\lambda)}\sum_{y,b} \Tr\left(B_{b|y}^{(\lambda)}\sigma_Z B_{b|y}^{(\lambda)} \left(U^{(\lambda)}_{by}\right)^\dagger C_{0|1}^{(\lambda)}U_{by}^{(\lambda)}\right).
\end{align}
We now examine separately the three types of projective measurement strategies,  (i-iii).

\begin{figure}
	\centering
	\includegraphics[width=\columnwidth]{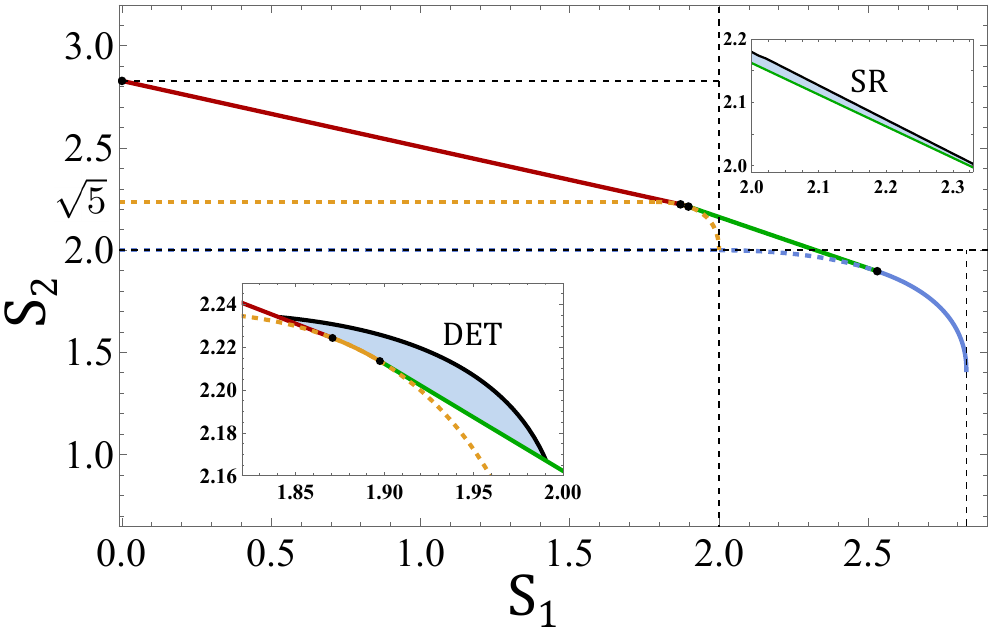}
	\caption{Optimal trade-off between $S_2$ and $S_1$ for a maximally entangled qubit pair under projective measurements and shared randomness. The trade-off consists of four parts represented by red, orange, green and blue solid curves respectively. The first and third are stochastic (boundary points marked) while the second and fourth are deterministic. The blue and orange dashed curves are the optimal trade-offs for deterministic projective strategies of type (i) and (iii) respectively. The lower (upper) inset illustrates, in solid black, a deterministic (stochastic) model based on partially entangled states that outperforms the maximally entangled state. The dashed black lines are the classical and quantum bounds of the CHSH inequality. }\label{FigRes}
\end{figure}

\textbf{Case (i):} ($\lambda=1$). Both measurements of $B_1$ are rank-1 projective. Let $A$ measure diagonally, by choosing $\theta=\frac{\pi}{4}$. Choose $B$'s observables as  $B_0=\cos\phi\sigma_X+\sin\phi\sigma_Z$ and $B_1=\sin\phi\sigma_X+\cos\phi\sigma_Z$. Then, we let the unitaries only depend on the input, i.e.~$U_{by}=U_y$, and choose $U_0=\openone$ and $U_1=e^{i\left(\phi-\frac{\pi}{4}\right)\sigma_Y}$. Finally, we choose $C$'s observables as $C_{0}=C_1=\cos\phi \sigma_X+\sin\phi\sigma_Z$. From \eqref{S1} we obtain $S_1^{(1)}=2\sqrt{2}\cos\phi$ and from \eqref{S2} we obtain $S_2^{(1)}=\sqrt{2}\left(\cos\phi+\sin\phi\right)$. The trade-off becomes 
\begin{equation}\label{rank1}
	S_2^{(1)}=\frac{1}{2}\left(S_1^{(1)}+\sqrt{8-\left(S_1^{(1)}\right)^2}\right),
\end{equation}
which we use in the range $2\leq S_1^{(1)}\leq 2\sqrt{2}$. In the remaining range, $0\leq S_1^{(1)}\leq 2$, one can straightforwardly choose observables to achieve the  trivial, classical, trade-off $S_2^{(1)}=2$. In Supplementary Material, we prove that there exists no strategy based on rank-one projective measurements that outperforms the above. 

\textbf{Case (ii):} ($\lambda=2$). Both measurements of $B$ are identity projections. This trivialises the trade-off. We write $B_0=B_1=\openone$. Rank-1 projective measurements for $A$ implies $S_1^{(2)}=0$. Optimally choosing $U_{by}=\openone$, the post-measurement state is left maximally entangled and hence we optimally choose $C_0=\sigma_X$ and $C_1=\sigma_Z$ and let $A$ measure diagonally to reach the Tsirelson bound $S_2^{(2)}=2\sqrt{2}$.

\textbf{Case (iii):} ($\lambda=3$). One measurement of $B$ is rank-1 projective and the other is an identity projection. We take setting $y=1$ as the former and outcome $b=0$ of setting $y=0$ to correspond to the projector $\openone$. This comes at no loss of generality because the CHSH parameter is invariant under the coordinated permutations $\{y\rightarrow \bar{y} \& a\rightarrow \bar{a} \text{ if } x=1\}$ and $\{b\rightarrow \bar{b} \text{ if } y=0 \& x\rightarrow \bar{x} \text{ and } a\rightarrow \bar{a}\}$ respectively, where the bar-sign denotes bitflip. Thus, we have $B_0=\openone$. Choose $B_1=\sigma_Z$, $U_{by}=\openone$, $C_0=\sigma_X$ and $C_1=\sigma_Z$. It yields $S_1^{(3)}=2\sin\theta$ and $S_2^{(3)}=\cos\theta+2\sin\theta$. The trade-off becomes
\begin{equation}\label{rankmixed}
	S_2^{(3)}=S_1^{(3)}+\frac{1}{2}\sqrt{4-\left(S_1^{(3)}\right)^2},
\end{equation}
which we use in the range $\frac{4}{\sqrt{5}}\leq S_1^{(3)}\leq 2$. Note that $S_1^{(3)}>2$ is not possible due to the trivialisation of setting $y=0$. The curve \eqref{rankmixed} has its maximum, $S_2^{(3)}=\sqrt{5}$, at $S_1^{(3)}=\frac{4}{\sqrt{5}}$. Therefore, in the range $0\leq S_1^{(3)}\leq \frac{4}{\sqrt{5}}$, one can easily find the trivial trade-off $S_2^{(3)}=\sqrt{5}$. In Supplementary Material, we prove that no better strategy exists for the given ranks of $B$. 

\textbf{Shared randomness:} We can now, via Eq.~\eqref{CHSH}, use the distribution of the shared randomness $\{p(\lambda)\}_{\lambda=1}^3$ to stochastically combine the three above cases to obtain the boundary of the region of attainable pairs $(S_1,S_2)$. This analysis is detailed in Supplementary Material. The boundary ends up being divided into four regions: a mixture between cases (ii)\&(iii), deterministic case (iii), a mixture between cases (i)\&(iii) and deterministic case (i). Specifically, the trade-off is given by  
\begin{align}\label{boundary}
	S_2=\begin{cases}
		\left(1-\frac{\sqrt{7}}{2}\right)S_1+2\sqrt{2} & \text{ if } 0\leq S_1 \leq \sqrt{\frac{7}{2}},\\
		S_1+\frac{1}{2}\sqrt{4-(S_1)^2} & \text{ if } \sqrt{\frac{7}{2}}\leq S_1\leq 3\sqrt{\frac{2}{5}},\\
		\sqrt{10}-\frac{S_1}{2} &  \text{ if }3\sqrt{\frac{2}{5}} \leq S_1 \leq 4\sqrt{\frac{2}{5}},\\
		\frac{S_1}{2}+\frac{1}{2}\sqrt{8-\left(S_1\right)^2} & \text{ if }4\sqrt{\frac{2}{5}}\leq S_1\leq 2\sqrt{2}.
	\end{cases}
\end{align}

This optimal trade-off is piecewise illustrated with solid lines in Figure~\ref{FigRes}. Crucially, we see that there exists a region, demarcated by the solid green line (correspondling to the third row of \eqref{boundary}), in which the CHSH inequality is violated twice. This proves that protocols based on L\"uders instruments, realising unsharp measurements, are not necessary for recycling nonlocality; projective measurements and just one bit of shared randomness is sufficient. In particular, the largest achievable identical double violation is $S_1=S_2=\frac{2\sqrt{10}}{3}\approx 2.108$. Notice also that all extremal  double violations can be realised with the same pair of quantum strategies, just by changing the distribution of the bit of shared randomness.

\textit{Stronger violations with partially entangled states.---} Consider that we substitute the maximally entangled state for a pure partially entangled state in the sequential CHSH scenario. These states take the form $\ket{\psi_\varphi}=\cos\varphi\ket{00}+\sin\varphi\ket{11}$, for some angle $\varphi\in[0,\frac{\pi}{4}]$. In the standard CHSH scenario, the strongest quantum nonlocality necessitates the maximally entangled state. Interestingly, in the sequential CHSH scenario based on projective measurements and shared randomness, the analogous turns out not to hold. We have numerically searched over quantum models consisting in stochastic mixtures of  type (i) and type (iii) strategies. For suitable choices of $\varphi$, we achieve double violations that are stronger than what is possible based on the maximally entangled state. In other words, partially entangled states enable points in the $(S_1,S_2)>(2,2)$ part of the $(S_1,S_2)$-plane that lie outside the boundary determined in Eq.~\eqref{boundary}. The numerics are illustrated in the upper inset in Figure~\ref{FigRes}. These strategies appear to be less elegant and less friendly to reproduce analytically than that detailed above for the maximally entangled state.

In Supplementary Material, we furthermore show that every entangled $\ket{\psi_\varphi}$ enables a double violations by mixing type (i) and type (ii) strategies. Moreover, we show that there exists deterministic type  (ii) and type (iii) strategies that for a suitable choice of $\varphi$ give rise to pairs of CHSH parameters $(S_1,S_2)$ that lie outside the boundary described in \eqref{boundary}. An example is illustrated in the lower inset of Figure~\ref{FigRes}. Nevertheless, these strategies cannot produce double violations due to their deterministic nature.

\textit{Double violations using only local classical randomness.---} The above protocols are based on active exploitation of classical shared randomness. A natural question is whether one can still recycle nonlocality, using only projective measurements, when the parties are classically independent, i.e.~without shared randomness. In this setting, classical randomness can only be locally generated by each party. For $n=2$, this amounts to replacing the collective classical variable $\lambda$ with a triple $(\lambda_A,\lambda_B,\lambda_C)$ whose elements are associated to $A$, $B$ and $C$ respectively. The independence corresponds to a factorising distribution $p(\lambda)=p(\lambda_A)p(\lambda_B)p(\lambda_C)$. From a resource point of view, this captures the most basic quantum approach to the sequential nonlocality scenario. 

Surprisingly, if $B$ leverages local randomness, it is still possible to achieve a double violation. We show it through an explicit quantum strategy. Let $A$ measure $A_0=\frac{\sqrt{3}}{2}\sigma_X+\frac{1}{2}\sigma_Z$ and $A_1=\cos(2)\sigma_X-\sin(2)\sigma_Z$, and let $C$ measure $C_0=\cos\left(\frac{2\pi}{3e}\right)\sigma_X-\sin\left(\frac{2\pi}{3e}\right)\sigma_Z$ and $C_1=\cos\left(\frac{1}{3}\right)\sigma_X+\sin\left(\frac{1}{3}\right)\sigma_Z$. Party $B$ randomly chooses between two strategies, labelled $\lambda_B\in\{0,1\}$, with some probability $q=p(\lambda_B=0)$. When $\lambda_B=0$, $B$ measures $B_0^{(0)}=\cos\left(\frac{2}{17}\right)\sigma_X+\sin\left(\frac{2}{17}\right)\sigma_Z$ and $B_1^{(0)}=\frac{\sigma_X+\sigma_Z}{\sqrt{2}}$ and applies unitaries $U^{(0)}_{00}=U^{(0)}_{10}=\openone$ and $U^{(0)}_{01}=U^{(0)}_{11}=e^{-\frac{2\pi}{27}i \sigma_Y}$. When $\lambda_B=1$, $B$ measures $B_0^{(1)}=\openone$ and $B_1^{(1)}=\frac{\sigma_X+\sigma_Z}{\sqrt{2}}$, and applies unitaries $U^{(1)}_{00}=e^{-\frac{5\pi}{81}i \sigma_Y}$, $U^{(1)}_{10}=\openone$ and $U^{(1)}_{01}=U^{(1)}_{11}=e^{-\frac{2\pi}{27}i \sigma_Y}$. This strategy is tailored for equal Bell parameters, which is achieved by choosing $q\approx 0.358$. It yields the double violation $S_1=S_2\approx 2.046$. We have also numerically searched for quantum strategies based on local randomness, but we have only found an improvement on the reported value on the sixth decimal.

\textit{Experimental considerations.---} The relevance of projective measurements for recycling nonlocality is not only interesting from a conceptual point of view. It also has noteworthy implications for experiments, which we now proceed to discuss. 

Firstly, our characterisation of the optimal trade-off \eqref{boundary} enables one to detect the use of quantum instruments that do not admit a realisation in terms of projective qubit measurements and qubit unitary operations. Natural examples are the more general, extremal, quantum instruments that correspond to unsharp measurements. Such certification is experimentally demonstrated in Ref.~\cite{Fan2022}. 
	
Secondly, the sufficiency of projective measurements enables one to demonstrate sequential nonlocality in much simpler setups than previously known. Standard protocols, based on L\"uders quantum instruments, entail entangling the qubit with an auxiliary degree of freedom, on which a projective measurement is performed. Such auxiliary entanglement requires more sophisticated  setups (see e.g.~\cite{Gomez2016, Smania2020, Smania2020b, Martinez2022}). In comparison, projective measurements and unitary transformations on single qubits are simpler, and the distribution of shared classical randomness (if used at all) requires no quantum technology. It may be noted that the violation magnitudes under projective measurements are smaller, $S_1=S_2\approx 2.108$, as compared to those reachable with unsharp measurements, $S_1=S_2\approx 2.263$ \cite{Silva2015}. However, if we for instance model experimental noise as an isotropic state $v\ketbra{\phi^+}{\phi^+}+\frac{1-v}{4}\openone$, then the projective double violation can be witnessed whenever $v\gtrsim 94.9\%$. Given the simplicity of the required quantum instrument, this is well within current capabilities in photonic experiments.

Thirdly, in photonics, which is the most relevant platform, a conceptually faithful implementation typically requires the introduction of additional photons for every unsharp measurement in the sequence, so that the original photon is not demolished in $B_k$'s measurement. Since this is challenging, it many times gives way for simplified, proof-of-principle, implementations where all outcomes of parties $B_1,\ldots, B_n$ are read out only in the final measurement of $B_n$. Such drawbacks are straightforwardly circumvented thanks to the simplicity of projective implementations.  For identity measurement, one does not need interact with the photon at all. For basis projections, the local post-measurement state is independent of the pre-measurement state. Hence, $B_k$ needs only to perform a standard measurement (demolishing the photon) and then prepare a new photon in the post-measurement state and relay it to $B_{k+1}$.

\textit{Outlook.---}  Contrasting the common wisdom, we have here shown that projective measurements are a resource for recycling quantum nonlocality. We have shown this for the most relevant case, namely that in which observers are independent, perform unbiased measurements and share a two-qubit entangled state. Given that such measurements also have been found useful for recycling quantum communication advantages \cite{Miklin2020}, it appears plausible that many other sequential quantum information protocols, e.g.~for entanglement witnessing \cite{Bera2018}, steering \cite{Sasmal2018} and contextuality \cite{Anwer2021}, also can be based on projective measurements. 

Our work leaves several relevant open problems. (1) It is known that for any $n$ there exists a protocol based on unsharp measurements on one share of a maximally entangled state that produces $n$ sequential violations of the CHSH inequality \cite{Brown2020}. How many sequential violations are possible under projective measurements? In Supplementary Material, we present a protocol that achieves $n=3$ sequential violations by exploiting shared randomness. However, we failed to numerically find a protocol for $n=4$. (2) We found that pure partially entangled states can produce larger double violations of the CHSH inequality than maximally entangled states.  What is  the optimal trade-off between the two CHSH parameters for a given partially entanged state $\ket{\psi_\varphi}$? In particular, what is the optimal trade-off when evaluated over all two-qubit entangled states? (3) There is considerable evidence \cite{Cheng2021, Chengb2021} that double violations of the CHSH inequality, under general measurements, are not possible when both qubits in the entangled pair are recycled. Could the introduction of shared randomness be interesting for this problem?

\begin{acknowledgements}
This work was supported by the Wenner-Gren Foundations. A.~S. acknowledges  the hospitality of the QUIT physics group.
\end{acknowledgements}

\bibliography{references_projective_sequence}


\onecolumngrid

\subsection*{Appendix A: Optimal trade-off for the maximally entangled state under projective measurements}\label{AppRegion}

\renewcommand\theequation{A\arabic{equation}}
\setcounter{equation}{0}
Consider that $A$ and $B$ share the maximally entangled state $\ket{\psi}=\ket{\phi^+}=\frac{1}{\sqrt{2}}\left(\ket{00}+\ket{11}\right)$. The optimal measurements of $A$ are rank-1 projective (these are extremal), because the qubit is not recycled afterwards. This amounts to standard observables $A_0=\vec{a}_0\cdot \vec{\sigma}$ and $A_1=\vec{a}_1\cdot \vec{\sigma}$ with $|\vec{a}_x|=1$. W.l.g.~we can take $\vec{a}_0=(\cos\theta,0,\sin\theta)$ and $\vec{a}_1=(\cos\theta,0,-\sin\theta)$ because any global rotation of $\vec{a}_0$ and $\vec{a}_1$ can be absorbed into a global rotation of $B$'s measurements via the relation $O\otimes \openone\ket{\phi^+}=\openone\otimes O^T\ket{\phi^+}$. 
 Thus, the unnormalised states remotely prepared by $A$ on $B$'s side correspond to the eigenvectors of the observables, namely $\rho_{a|x}=\frac{1}{4}\left(\openone+(-1)^a\vec{a}_x\cdot \vec{\sigma}\right)$. Here, $p(a|x)=\Tr\left(\rho_{a|x}\right)=\frac{1}{2}$ which follows from the fact that $A$'s measurement operators are trace-one and the local state is maximally mixed. We can then define $\rho_x=\rho_{0|x}-\rho_{1|x}=\frac{\vec{a}_x\cdot\vec{\sigma}}{2}$. The CHSH parameter between $A$ and $B$ now reads
\begin{equation}\label{chsh1app}
S_{1}=\sum_{x,y=0,1}(-1)^{xy}\Tr\left(\rho_{x}B_y\right),
\end{equation}
where $B_y$ is $B$'s observable. 

Every quantum instrument that realises a measurement $\{B_{0|y},B_{1|y}\}$  can be represented as a square-root instrument followed by an arbitrary CPTP map \cite{Brown2020}. The square-root instrument performs on the input state the transformation $\rho\rightarrow \sqrt{B_{b|y}}\rho \sqrt{B_{b|y}}$. Since extremal CPTP maps taking a qubit to a qubit are unitary, we can write the Kraus operators of the quantum instrument as $K_{b|y}=U_{by}\sqrt{B_{b|y}}$. Hence, state shared between $A$ and $C$ after $B$'s measurement reads
\begin{equation}\label{post}
\psi_2= \frac{1}{2}\sum_{y,b} \left(\openone\otimes U_{by}\sqrt{B_{b|y}}\right)\psi \left(\openone\otimes \sqrt{B_{b|y}} U^\dagger_{by}\right).
\end{equation}
Similarly, the CHSH parameter between $A$ and $C$ becomes
\begin{align}\nonumber\label{chsh2app}
S_{2}&=\frac{1}{2}\sum_{x,z=0,1}(-1)^{xz}\sum_{y,b}\Tr\left(U_{by}\sqrt{B_{b|y}}\rho_{x}\sqrt{B_{b|y}} U^\dagger_{by}C_{0|z}\right)\\
	&=\cos\theta\sum_{y,b} \Tr\left(\sqrt{B_{b|y}}\sigma_X\sqrt{B_{b|y}} U^\dagger_{by}C_{0|0}U_{by}\right)+\sin\theta\sum_{y,b} \Tr\left(\sqrt{B_{b|y}}\sigma_Z\sqrt{B_{b|y}} U^\dagger_{by}C_{0|1}U_{by}\right),
\end{align} 
where we for simplicity have used normalisation to make the substitution $C_z=2C_{0|z}-\openone$ and used the cyclicity of the trace. We can without loss of generality restrict $C$'s measurements to be rank-1 projective. 

We now proceed to examine the trade-off between $S_{1}$ and $S_{2}$ when $B$'s measurements are restricted to being projective, i.e.~$B_{b|y}B_{b'|y}=\delta_{b,b'}B_{b|y}$, causing the relevant Kraus operators take the form $K_{b|y}=U_{by}B_{b|y}$. Thus, the second CHSH parameter simplifies into
\begin{align}\label{S2A}
S_{2}=\cos\theta\sum_{y,b} \Tr\left(B_{b|y}\sigma_X B_{b|y} U^\dagger_{by}C_{0|0}U_{by}\right)+\sin\theta\sum_{y,b} \Tr\left(B_{b|y}\sigma_Z B_{b|y} U^\dagger_{by}C_{0|1}U_{by}\right).
\end{align} 
Classifying the projectors by their rank, there are only three qualitatively different cases.
\begin{enumerate}[(i)]
	\item Both $B$'s measurements are projections onto two orthogonal vectors (standard qubit measurement).
	\item Both $B$'s measurements are projections onto the identity and the zero projector respectively (trivial measurement).
	\item One of $B$'s measurements is a projection onto two orthogonal vectors and the other is trivial.
\end{enumerate}
In what follows, we examine these three cases one by one.

\subsection*{1. Case (i): projection onto a basis}\label{AppRegion1}
We now characterise the optimal trade-off between $S_{1}$ and $S_{2}$ when $B$ is restricted to performing rank-one projective measurements. Note that such strategies enable $S_{1}>2$ but not $S_{2}>2$ because the instrument of $B$ is entanglement breaking. 

Define the rank-one projectors $P_{zyb}=U^\dagger_{by}C_{0|z}U_{by}$. We can place an upper bound on $S_{2}$ by assuming that $P_{zyb}$ can be aligned with the eigenvector of $B_{b|y}\sigma_X B_{b|y}$ (when $z=0$) and the eigenvector of $B_{b|y}\sigma_Z B_{b|y}$ (when $z=1$), both  associated to the largest eigenvalue. This gives the upper bound
\begin{equation}\label{ll}
S_{2}\leq \cos\theta\sum_{y,b} \lambda_\text{max}\left(B_{b|y}\sigma_X B_{b|y} \right)+\sin\theta\sum_{y,b} \lambda_\text{max}\left(B_{b|y}\sigma_Z B_{b|y}\right).
\end{equation}
Notice now that operators of the form $P(\vec{u}\cdot\vec{\sigma})P$, where $P$ is a rank-one projector, are rank-one. Hence, their spectra is of the form $(0,\lambda)$. We may write $\lambda_\text{max}\left(P(\vec{u}\cdot\vec{\sigma})P\right)=\max\{0,\Tr\left((\vec{u}\cdot\vec{\sigma})P\right)\}$. Since the second argument is just the expectation value of measuring the state $P$ with the observable $\vec{u}\cdot\vec{\sigma}$, it follows that for each $y$ and each of the two terms in \eqref{ll}, we get a contribution from only one value of $b$. Moreover, since the states remotely prepared by $A$ for $B$ are in the XZ-plane, it is optimal to assign Bloch vectors for $B$ in the same plane:  $\vec{b}_y=\left(\cos\phi_y,0,\sin\phi_y\right)$. Then, one finds
\begin{equation}
S_{2}\leq \cos \theta \left(|\cos\phi_0|+|\cos\phi_1|\right)+\sin \theta \left(|\sin\phi_0|+|\sin\phi_1|\right).
\end{equation}
We may also write the CHSH parameter between $A$ and $B$ as
\begin{equation}
S_{1}=(\vec{a}_0+\vec{a}_1)\cdot \vec{b}_0+(\vec{a}_0-\vec{a}_1)\cdot \vec{b}_1=2\cos\theta\cos\phi_0+2\sin\theta\sin\phi_1.
\end{equation}
Thus, the pair $(S_{1},S_{2})$ is characterised by the variables $(\theta,\phi_0,\phi_1)$. 

Firstly, we note that by choosing $\phi_0=\phi_1=-\theta$, we obtain $S_1=2\cos\left(2\theta\right)$ and $S_2=2$. By taking $\theta\in[0,\frac{\pi}{4}]$, we cover the classical boundary, namely $S_2(S_1)=2$ for $0\leq S_1\leq 2$. This is optimal because a rank-1 projective strategy leaves the state between $A-C$ separable and hence it can at best saturate the CHSH inequality.

Secondly, we consider the non-trivial range $2\leq S_1\leq 2\sqrt{2}$. We claim that for every pair $(S_{1},S_{2}')$ obtained at $(\theta,\phi_0,\phi_1)$, there exists another pair $(S_{1},S_{2})$, with $S_{2}\geq S_{2}'$, obtained at $(\theta=\frac{\pi}{4}, \phi_0=\phi,\phi_1=\frac{\pi}{2}-\phi)$ for some $\phi\in[0,\frac{\pi}{2}]$. To show this, we must first reproduce $S_{1}$ with the new strategy. This implies that we must find a $\phi$ such that
\begin{equation}\label{phistep}
2\cos\theta\cos\phi_0+2\sin\theta\sin\phi_1=2\sqrt{2}\cos\phi.
\end{equation}
The left-hand-side is maximal for $\phi_0=0$, $\phi_1=\frac{\pi}{2}$ and $\theta=\frac{\pi}{4}$, for which it becomes $2\sqrt{2}$. Thus, for any value of the left-hand-side we can always choose $\phi$ such that equality holds. Next, we show that with this choice of $\phi$, we have $S_{2}\geq S_{2}'$. This condition becomes
\begin{equation}
\cos \theta \left(|\cos\phi_0|+|\cos\phi_1|\right)+\sin \theta \left(|\sin\phi_0|+|\sin\phi_1|\right)\leq \sqrt{2}\left(\cos\phi+\sin\phi\right).
\end{equation}
In the range of our interest, we can w.l.g.~take $\phi_0,\phi_1\in[0,\frac{\pi}{2}]$ and drop the absolute values. Squaring both sides and using \eqref{phistep} to substitute for $\phi$, we arrive at 
\begin{equation}\label{phistep2}
\cos^2\theta \left(\cos^2\phi_0+\cos^2\phi_1\right)+\sin^2\theta \left(\sin^2\phi_0+\sin^2\phi_1\right)+\sin(2\theta)\sin\left(\phi_0+\phi_1\right)\leq 2.
\end{equation}
Differentiating the left-hand-side w.r.t. $\phi_0$ and $\phi_1$ respectively, we find that they have two joint roots, at $\theta=\frac{\pi}{4}$, $\phi_0+\phi_1=\frac{\pi}{2}$, and at $\theta=\phi_0=\phi_1$. In both cases, the derivative w.r.t. $\theta$ vanishes. Inserting this into the left-hand-side of \eqref{phistep2}, we find that the maximum is $2$, thus proving the inequality to hold.

Hence, we need only to consider strategies of the form $\theta=\frac{\pi}{4}$, $\phi_0=\phi$ and $\phi_1=\frac{\pi}{2}-\phi$. This gives
\begin{align}
&S_{1}=2\sqrt{2}\cos\phi \\
&S_{2}\leq \sqrt{2}\left(\cos\phi+\sin\phi\right).
\end{align}
Substituting the former into the latter, we see that the trade-off between the two is given by
\begin{equation}\label{i}
S_{2}(S_1)\leq \frac{S_1}{2}+\frac{1}{2}\sqrt{8-(S_1)^2}.
\end{equation}
This is optimal (tight) because it equals the trade-off obtained from an explicit quantum strategy in the main text. Moreover, notice that the maximum of this function occurs at $S_1=2$, where we have $S_2=2$. Thus, at the lower demarcation of its range of validity, $2\leq S_1\leq 2\sqrt{2}$, it meets the trivial (classical) trade-off encountered for $0\leq S_1\leq 2$.

\subsection*{2. Case (ii): identity projections}\label{AppRegion2}
The trade-off between $S_{1}$ and $S_{2}$ when $B$ performs trivial projective measurements, corresponding to deterministically choosing $b$ based on $y$ without regard to the quantum state, is trivial. Such measurements are represented as either $(\openone,0)$ (always output $b=0$) or $(0,\openone)$ (always output $b=1$). From \eqref{chsh1app} it follows that $S_{1}=0$. Notice that this can be increased to $S_{1}=2$ if $A$ also performs trivial measurements, but this is not interesting in our context because it implies $S_{2}\leq 2$. Thus, $B$'s instrument is essentially reduced only to a unitary: the post-measurement state becomes $\tilde{\phi}^+=\frac{1}{2}\sum_{y} \left(\openone\otimes V_y\right) \phi^+ \left(\openone\otimes V^\dagger_{y}\right)$ where $V_y$ is the element in $\{U_{0y},U_{1y}\}$ associated to the single output (unit probability event) of $B$'s measurements. Clearly, one optimal choice is $V_y=\openone$, for which $\tilde{\phi}^+=\phi^+$. Then, $C$ can achieve the Tsirelson bound $S_{2}=2\sqrt{2}$ by performing measurements $\sigma_X$ and $\sigma_Z$, while $A$ chooses $\theta=\frac{\pi}{4}$. In contrast to case (i) and case (iii), this trivial trade-off is just a single point: $(S_1,S_2)=(0,2\sqrt{2})$.

\subsection*{3. Case (iii): one identity projection and one basis projection}\label{AppRegion3}
Recall from the main text that we can w.l.g.~choose to associate $B$'s identity measurement to the setting $y=0$ and the single relevant outcome to $b=0$. Thus, we have $B_0=\openone$. The second measurement is a standard basis projection, corresponding to measuring in the direction of the unit Bloch vector $\vec{b}$. The observable is $B_1=\vec{b}\cdot \vec{\sigma}$.

Due to the freedom of a global shift in the unitaries $U_{by}$, we may always fix the first one as a reference $U_{00}=\openone$. Moreover, since $B_{1|0}=0$, the choice of $U_{10}$ does not influence the post-measurement state \eqref{post}. Furthermore, for $y=1$, recall from case (i) that the eigenvector of $B_{b|y}(\vec{u}\cdot\vec{\sigma}) B_{b|y}$, for any unit vector $\vec{u}$, associated to its largest eigenvalue, is identical for $b=0$ and $b=1$ (one eigenvalue is positive, the other is zero). In Eq.\eqref{S2A}, the optimal unitaries $U_{b1}$ aim to align both $C_{0|0}$ and $C_{0|1}$ with said eigenvector. Since this vector does not depend on $b$, we optimally choose $U_{01}=U_{11}\equiv U_1$. The post-measurement state becomes 
\begin{equation}
\frac{1}{2}\phi^++\frac{1}{4}\left(\openone\otimes U_1\right) \left(\ketbra{\vec{b},\vec{b}}{\vec{b},\vec{b}}+\ketbra{-\vec{b},-\vec{b}}{-\vec{b},-\vec{b}}\right)\left(\openone\otimes U_1^\dagger\right) .
\end{equation}

Since the states prepared remotely by $A$ for $B$ are in the XZ-plane, it is optimal to choose $\vec{b}=\left(\cos\phi,\sin\phi\right)$. This gives $S_{1}=2\sin\theta\sin\phi$. The only remaining unitary is then optimally taken as a rotation in the same plane, $U_1=e^{i\mu \sigma_Y}$. 
Writing $C$'s measurement Bloch vectors as $\vec{c}_z=(\cos\phi_z,0,\sin\phi_z)$, we have
\begin{equation}
S_{2}=\frac{1}{2}\cos\theta\left(\cos(2\mu+2\phi-\phi_0)+\cos(2\mu-\phi_0)+2\cos\phi_0\right)+\frac{1}{2}\sin\theta\left(\sin(2\mu+2\phi-\phi_1)-\sin(2\mu-\phi_1)+2\sin\phi_1\right).
\end{equation}

Considering the derivative w.r.t.~$\mu$, $\phi_0$, $\phi_1$, $\phi$ respectively, one finds that they are  simultaneously zero when $\phi_0=\mu=0$ and $\phi=\phi_1=\frac{\pi}{2}$.
This corresponds to the optimum, at which we obtain 
\begin{equation}
S_{2}=\cos\theta+2\sin\theta.
\end{equation}
Note that $\phi=\frac{\pi}{2}$ also optimises $S_1$, yielding $S_1=2\sin\theta$. Thus, the optimal trade-off is given by
\begin{equation}\label{iii}
S_{2}=S_1+\sqrt{1-\frac{(S_1)^2}{4}}.
\end{equation}
This reaches its maximal value at $S_1=\frac{4}{\sqrt{5}}$ where we have $S_{2}=\sqrt{5}>2$. In the range $0\leq S_1\leq \frac{4}{\sqrt{5}}$, the above solution is no longer optimal. However, in this range the optimal trade-off is trivially given by $S_2(S_1)=\sqrt{5}$ because the largest value of $S_2$, when no consideration is given to $S_1$ other than fixing the ranks of $B$'s measurements, is $\sqrt{5}$. This is seen from the fact that choosing $\phi_0=0$, $\phi=\phi_1=\frac{\pi}{2}$ and $\mu=\frac{1}{2} \arccos\left(\frac{\sqrt{5}-\sin\theta}{\sin \theta+\cos\theta}\right)$ gives $S_2=\sqrt{5}$ and simultaneously leaves $S_1$ free to change.

\subsection*{4. Mixing via shared randomness}\label{AppRegionMix}
We deduce the optimal trade-off between $S_{1}$ and $S_{2}$ by stochastically combining the projective strategies using shared randomness. Recall that case (i) enables a first violation but not a second violation, whereas cases (ii) and (iii) do not enable a first violation but do enable a second violation. Notice that in all three deterministic cases, the trade-offs are concave. Thus, it is only useful to leverage shared randomness in order to mix between deterministic strategies associated to different rank combinations (i-iii). We consider the mixing case by case.

\textbf{Mixing (ii) and (iii):} Define $f(x)=x+\frac{1}{2}\sqrt{4-x^2}$, corresponding to the function form of the optimal trade-off \eqref{iii} for strategies of class (iii). We have $\frac{df}{dx}=1-\frac{x}{2\sqrt{4-x^2}}$. The tangent of $f$ is then obtained from solving 
\begin{equation}
\left(1-\frac{x}{2\sqrt{4-x^2}}\right)x+k=f(x) \Rightarrow k=\frac{2}{\sqrt{4-x^2}}.
\end{equation}
We are seeking the tangent that intersects the point $(0,2\sqrt{2})$ associated to case (ii), which means putting $k=2\sqrt{2}$. This gives $x=\sqrt{\frac{7}{2}}$. Thus, one piece of the boundary in the space of $(S_1,S_2)$ becomes
\begin{equation}
S_2=\left(1-\frac{\sqrt{7}}{2}\right)S_1+2\sqrt{2},
\end{equation}
which is valid for $0\leq S_1\leq \sqrt{\frac{7}{2}}$.

\textbf{Mixing (i) and (iii):} Consider now a mixture of the two non-trivial strategies. Thus, we look for the common tangent connecting the curves \eqref{i} and \eqref{iii}. For the former we write $g(x)=\frac{x}{2}+\frac{1}{2}\sqrt{8-x^2}$ and for the latter we write $f(x)=x+\sqrt{1-\frac{x^2}{4}}$. To find the tangent, we solve the two equations
\begin{equation}
	f'(x_1)=g'(x_2)=\frac{f(x_1)-g(x_2)}{x_1-x_2},
\end{equation}
where $f'$ and $g'$ are the derivatives of $f$ and $g$. The solution is 
\begin{align}
	& x_1=3\sqrt{\frac{2}{5}} & x_2=4\sqrt{\frac{2}{5}}.
\end{align}
Returning to our original notations, this gives the tangent 
\begin{equation}\label{part3}
	S_2=\sqrt{10}-\frac{S_1}{2},
\end{equation}
which constitutes another piece of the boundary in the space of $(S_1,S_2)$, valid in the interval $3\sqrt{\frac{2}{5}}\leq S_1\leq 4\sqrt{\frac{2}{5}}$.

\textbf{Mixing (i) and (ii):} Computing the tangent between the point $(0,2\sqrt{2})$ and the curve \eqref{i} is straightforward. However, it turns out to not correspond to the boundary: one can achieve a better trade-off by mixing other strategies, e.g.~as in the above cases. Thus, this case is not interesting.

\textbf{Intermediate regions:} We are left to determine the optimal trade-off for the remaining two parts of the interval $0\leq S_1\leq 2\sqrt{2}$. These are
\begin{align}\label{s1}
&\sqrt{\frac{7}{2}}\leq S_1\leq 3\sqrt{\frac{2}{5}},\\\label{s2A2}
& 4\sqrt{\frac{2}{5}}\leq S_1\leq 2\sqrt{2}.
\end{align}
Since these are not covered by the first two cases (considered above), and the mixture of (i) and (ii) is suboptimal, we must attribute them to deterministic strategies. Specifically, in the first part, namely in the interval \eqref{s1}, the deterministic strategy (iii) is optimal (see Eq.~\eqref{iii}). In the second part, namely in the interval \eqref{s2A2}, the deterministic strategy (i) is optimal (see Eq.~\eqref{i}). Thus, the full boundary of the set of $(S_1,S_2)$ reachable under projective measurements and shared randomness is given by a four-part piecewise function.

\subsection*{Appendix B: Sequential scenario with partially entangled states}\label{AppPartial}

\renewcommand\theequation{B\arabic{equation}}
\setcounter{equation}{0}

\begin{figure}
	\centering
	\includegraphics[width=0.6\columnwidth]{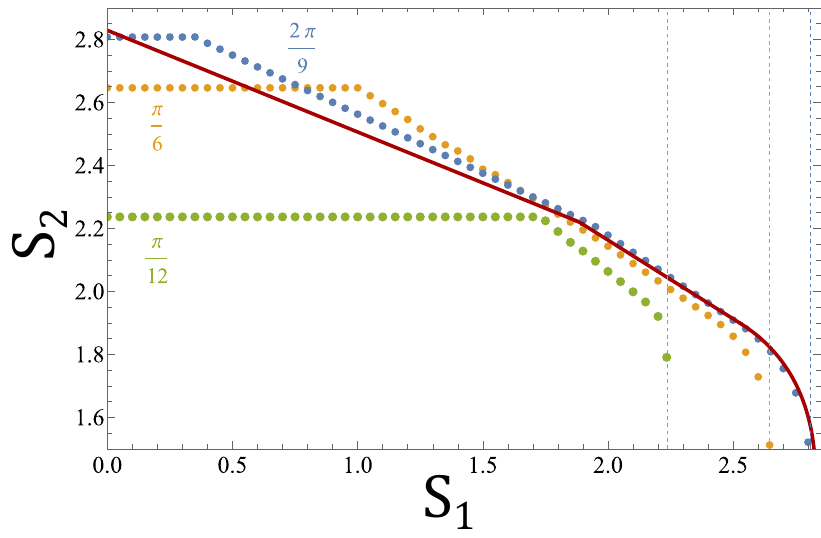}
	\caption{Numerical estimates for the optimal trade-off between $S_1$ and $S_2$ for the partially entangled state $\ket{\phi_\varphi}$. The solid red line represents the analytically obtained optimal trade-off for the maximally entangled state. The dashed lines represent the optimal violations of the CHSH inequality enabled by $\ket{\psi_\varphi}$.}\label{AppFig}
\end{figure}

Consider that the shared state is partially entangled, $\ket{\psi_\varphi}=\cos\varphi\ket{00}+\sin\varphi\ket{11}$, for some $\varphi\in[0,\frac{\pi}{4}]$. Party $A$'s and party $C$'s measurements can w.l.g.~be restricted to the XZ-plane. Similarly we can restrict those measurements of $B$ that are rank-1 to the XZ-disk as well. We have investigated the optimal value of $S_2$ for a given value of $S_1$ by numerically optimising the former over the measurements of $A$, $B$ and $C$, as well as the unitaries of $B$ (which may be restricted to the $XZ$-plane) and the distribution $\{p(\lambda)\}_{\lambda=1}^3$ of a trit of shared randomness. The results for a few selected values of $\varphi$ are illustrated in Figure~\ref{AppFig}. Notice that we find many points that go beyond the optimal boundary determined analytically in the main text for the maximally entangled state. In particular,  for $\varphi=\frac{2\pi}{9}$, we find instances of improved double violations.

We present a simple analytical strategy for case (i), i.e.~rank-1 projective measurements for $B$. Via numerics, we have found no better deterministic type (i) strategy for any value of $\varphi$.  Party $A$ measures $A_0=\sigma_X$ and $A_1=\sigma_Z$. Party $B$ measures $B_0=\cos\phi\sigma_X+\sin\phi\sigma_Z$ and $B_1=\cos\phi\sigma_X-\sin\phi\sigma_Z$, and then applies the unitaries $U_{by}=U_y$ with $U_0=\openone$ and $U_1=e^{i\left(\phi-\frac{\pi}{2}\right)\sigma_Y}$. Party $C$ measures $C_0=\cos\phi\sigma_X+\sin\phi\sigma_Z$ and $C_1=-\cos\phi\sigma_X-\sin\phi\sigma_Z$. This gives 
\begin{align}
&S_{1}=2\left(\cos\phi\sin(2\varphi)+\sin\phi\right)\\
&S_{2}=2\sin\phi.
\end{align}
Using $\cos\phi=\sqrt{1-\sin^2\phi}$, solving the first equation for $\sin\phi$, and substituting into the second equation gives  
\begin{equation}\label{tt}
S_{2}=\frac{1}{1+\sin^2(2\varphi)}\left(S_1+\sin\left(2\varphi\right)\sqrt{4\left(1+\sin^2\left(2\varphi\right)\right)-(S_1)^2}\right)
\end{equation}
Note that  this reduces to \eqref{i} for $\varphi=\frac{\pi}{4}$.

\subsection*{1. Deterministic type (ii) strategy}
Furthermore, we may note that one can go beyond the maximally entangled state using only type (ii) deterministic strategies for partially entangled states. To see this, simply note that by choosing $B_0=B_1=\openone$, the CHSH parameter becomes $S_1=2\expect{A_0\otimes \openone}=2\Tr\left(A_0\left(\cos^2\varphi\ketbra{0}{0}+\sin^2\varphi\ketbra{1}{1}\right)\right)$. The optimal rank-1 measurement for $A$ is $A_0=\sigma_Z$ which gives $S_1=2\cos\left(2\varphi\right)$. This is an improvement on the case of the maximally entangled state for which $S_1=0$. Then, we may note that for any state $\ket{\psi_\varphi}$, the optimal CHSH value is $2\sqrt{1+\sin^2\left(2\varphi\right)}$ \cite{Horodecki1995} and that this is achievable by fixing $A_0=\sigma_Z$. Thus, optimal type (ii) strategies achieve
\begin{align}\label{typeii}
&	S_1=2\cos\left(2\varphi\right), &S_2= 2\sqrt{1+\sin^2\left(2\varphi\right)}.
\end{align}
This can be compared to the line segment $S_2(S_1)=\left(1-\frac{\sqrt{7}}{2}\right)S_1+2\sqrt{2}$ of the boundary associated to the maximally entangled state. We have $\varphi=\frac{1}{2}\arccos\left(\frac{S_1}{2}\right)$ which gives $S_2=\sqrt{8-(S_1)^2}$ for the type (ii) strategy. It exceeds the line segment whenever $0<S_1<h$ where  
\begin{equation}
	h=\frac{8\sqrt{2}}{113}\left(7\sqrt{7}-2\right)\approx 1.65.
\end{equation}

\subsection*{2. Deterministic type (iii) strategy}
We show that one can go beyond the boundary in the $(S_1,S_2)$-plane for the maximally entangled state using deterministic type (iii) strategies and partially entangled states. Party $A$ measures $A_x=(-1)^x \cos\theta\sigma_X+\sin\theta\sigma_Z$ while party $B$ measures $B_0=\openone$ and $B_1=\sigma_X$, ignoring any subsequent unitary ($U_{by}=\openone$). Party $C$ measures $C_0=\sigma_Z$ and $C_1=\sigma_X$. A direct calculation gives $S_1=2\sin\left(\theta+2\varphi\right)$ and $S_2=\sin\theta+2\cos\theta\sin\left(2\varphi\right)$. The former implies $\theta=\pi-2\varphi-\arcsin\left(\frac{S_1}{2}\right)$, which for the latter implies
\begin{align}\nonumber
S_2=&\sin\left(2\varphi\right)\sqrt{1-\left(\frac{S_1}{2}\right)^2}\left(1-2\cos\left(2\varphi\right)\right)\\
&+\frac{S_1}{2}\left(2\sin^2\left(2\varphi\right)+\cos\left(2\varphi\right)\right).
\end{align}
The optimal choice of $\varphi$ is found to be 
\begin{equation}
\varphi=\arccos\left(\frac{1}{4}\sqrt{9-\sqrt{g(S_1)}+\sqrt{33-g(S_1)+\frac{8(S_1)^2}{\sqrt{g(S_1)}}}}\right),
\end{equation}
where $g(x)=11+h(x)+(121-24x^2)/h(x)$ and $h(x)=\left[8x^4-396 x ^2+8x^2 \sqrt{x^4+117 x^2-484} +1331\right]^{1/3}$. This outperforms the maximally entangled state in the range $1.84\lesssim S_1 \lesssim 1.99$ (see figure in main text). As $S_1$ increases, the entanglement becomes weaker, reaching about $\varphi\approx 0.686$ at the upper demarcation ($S_1\approx 1.99$). We have also numerically investigated deterministic, type (iii), strategies in this range but only found a tiny improvement in $S_2$ as compared to the above analytical construction: the improvement increases with $S_1$ and was at most found to be roughly of size $2\times 10^{-3}$.

\subsection*{3. All pure entangled states produce double violations}
Finally, we show that every pure entangled state $\ket{\psi_\varphi}$ can produce a double violation of the CHSH inequality. This is intuitively clear by inspecting the $(S_1,S_2)$-plane. Firstly, for $0<\varphi\leq \frac{\pi}{4}$, the type (ii) strategy \eqref{typeii} provides a point in the $(S_1,S_2)$-plane that has $S_1<2$ and $S_2>2$. As $\varphi$ decreases, this point approaches $(2,2)$. Secondly, the above type (i) strategy \eqref{tt} only depends on $S_1$ to second order in the vicinity of $S_1=2$. Putting $S_1=2+\epsilon$ for some small $\epsilon>0$, the Taylor expansion of \eqref{tt} gives
\begin{equation}\label{tailor}
S_2=2-\frac{\csc^2\left(2\varphi\right)}{4}\epsilon^2+O(\epsilon^3).
\end{equation}
Thus, for sufficiently small $\epsilon\approx 0$, the trade-off is essentially flat. Mixing these two strategies over shared randomness allows us to realise the secant that connects the point \eqref{typeii} with the curve \eqref{tailor}. Thus, by choosing an infinitesimal $\epsilon$ we assure the validity of the Taylor expansion, and necessitate that the secant passes through the double violation region $(S_1,S_2)>2$.

\begin{figure}
	\centering
	\includegraphics[width=0.6\columnwidth]{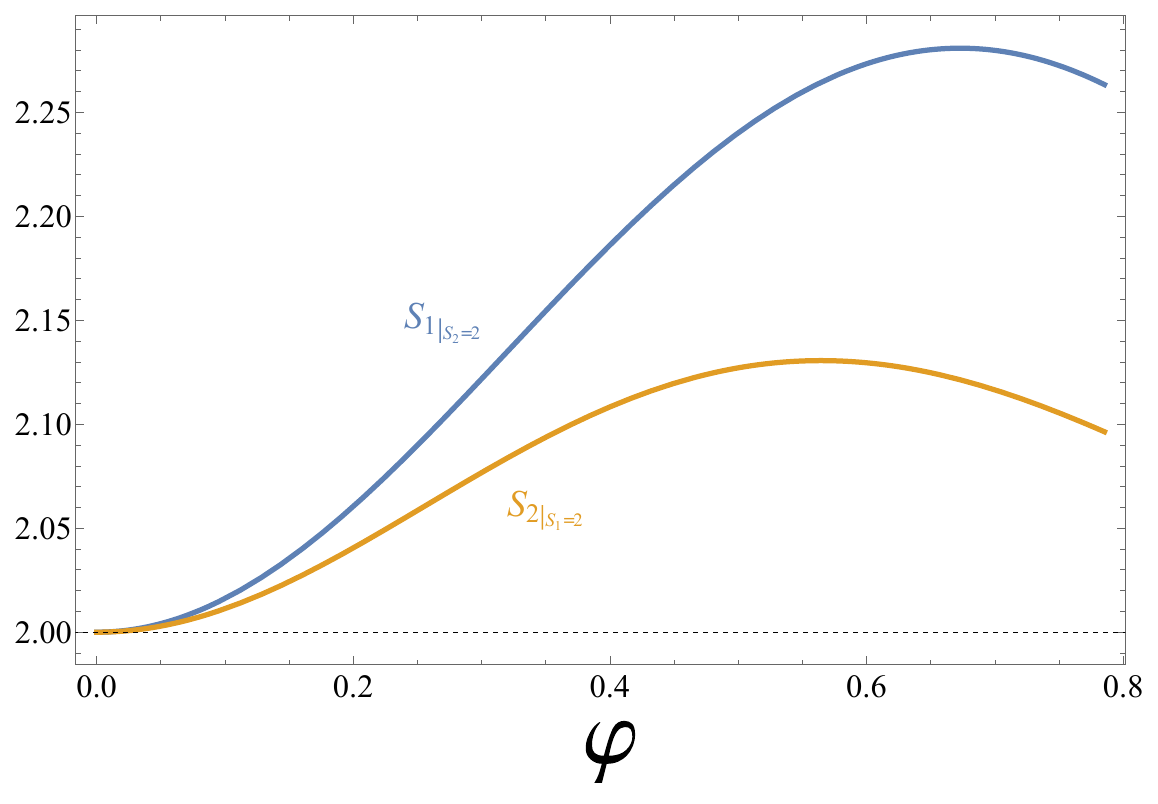}
	\caption{Dependendence of the points $S_2(S_1=2)$ and $S_1(S_2=2)$ on $\varphi$ for the tangent between the point \eqref{typeii} and the curve \eqref{tt}. Since both curves are always (except for $\varphi=0$) above the local bound, $2$,  the tangent passes through the double violation region $(S_1,S_2)>2$.}\label{FigPointEvolution}
\end{figure}

One can also find the result through direct calculation, e.g.~by evaluating the tangent between the point \eqref{typeii} and the curve \eqref{tt}, and then check the points at which this tangent intersects the line $S_2=2$ and $S_1=2$ respectively. The expression for these two points is cumbersome, but we plot it in Figure~\ref{FigPointEvolution}. Note that both curves exceed the local bound whenever $\varphi\neq 0$. For small $\varphi$, namely $\varphi\approx 0$, we have the tailor expansions
\begin{align}
	& S_1(S_2=2)=2+4\left(\sqrt{2}-1\right)\varphi^2\\
	& S_2(S_1=2)=2+2\left(2-\sqrt{2}\right)\varphi^2,
\end{align}
which demonstrate that both points are above the local bound. 

\subsection*{Appendix C: Double violation without unitaries in the quantum instrument}

\renewcommand\theequation{C\arabic{equation}}
\setcounter{equation}{0}
We give a simple example showing that a double violation also is possible based on a quantum instrument consisting in a projective measurement that is not followed by a unitary.Let the state be maximally entangled, $\ket{\psi}= \ket{\phi^+}$, and let the parties share one bit of randomness $\lambda\in\{1,2\}$.

When $\lambda=1$, the pair $A-B$ runs the standard quantum strategy for attaining the Tsirelson bound. Thus, $B$ measures $\sigma_X$ and $\sigma_Z$, and  $A$ measures diagonally, yielding $S_1^{(1)}=2\sqrt{2}$. Here $(\sigma_X,\sigma_Y,\sigma_Z)$ denote the standard Pauli matrices. We ignore the unitaries ($U_{by}^{(1)}=\openone$). Let $C$ also measure $\sigma_X$ and $\sigma_Z$. This gives $S_2^{(1)}=\sqrt{2}$. 

When $\lambda=2$, we set $B$'s first observable to $B_0=\openone$ and choose the second observable as a basis projection. Since the local state is $\frac{\openone}{2}$, we can w.l.g.~choose $B_1=\sigma_Z$. Again ignoring the unitaries  ($U_{by}^{(2)}=\openone$), the state of $A-C$ becomes $\frac{3}{4}\phi^++\frac{1}{4}\phi^-$. Via the Horodecki criterion \cite{Horodecki1995}, one finds that the state enables at most the CHSH value $S_2^{(2)}=\sqrt{5}$. The measurements that achieve this are  $\frac{1}{\sqrt{5}}\sigma_X\pm \frac{2}{\sqrt{5}}\sigma_Z$ for $A$, and $\sigma_X$ and $\sigma_Z$ for $C$. These choices also imply  $S_1^{(2)}=\frac{4}{\sqrt{5}}$.

Now, we use the shared randomness to mix the two projective measurement strategies together. Notice also that $\lambda$ only is used to correlate $A-B$, since $C$ performs the same measurements in both strategies. The final pair of CHSH parameters becomes 
\begin{align}
&S_1=q S_1^{(1)}+(1-q)S_1^{(2)}, \\
& S_2=q S_2^{(1)}+(1-q)S_2^{(2)},
\end{align}
where  $q=p(\lambda=1)$ is the degree of mixture between the two strategies.  If, for instance, we insist on both CHSH parameters being equal, we can immediately determine $q$ and obtain the double violation $S_1=S_2=\frac{6 \sqrt{10}}{5 \sqrt{2}+\sqrt{5}}\approx 2.04$.

\subsection*{Appendix D: Three sequential violations with projective measurements and shared randomness}

\renewcommand\theequation{D\arabic{equation}}
\setcounter{equation}{0}

We now go beyond the case of two sequential parties and show that the maximally entangled state $\ket{\phi^+}$ enables also a triple violation of the CHSH inequality. To show this, let all parties share a trit of randomness $\lambda\in\{1,2,3\}$. For each $\lambda$, we tailor a quantum strategy such that the pair $A-B_\lambda$ violates the CHSH inequality while the other two pairs only nearly fail to achieve a violation. 

For $\lambda=1$, let $A$ measure $A_{x}=\frac{\sigma_X+(-1)^x\sigma_Z}{\sqrt{2}}$, and choose  $B_0^{1}=\cos\phi \sigma_X+\sin\phi\sigma_Z$ and  $B_1^{1}=\sin\phi \sigma_X+\cos\phi\sigma_Z$ followed by unitaries $U_{b_1y_1}^{1}=U_{y_1}^{1}$, with $U_{0}^{1}=\openone$ and $U_{1}^{1}=e^{i(\phi-\frac{\pi}{4})\sigma_Y}$. The remaining two sequential parties perform no unitaries ($U=\openone$) and identical measurements: $B_{y}^{2}=B_{y}^{3}=\cos\phi\sigma_X+\sin\phi\sigma_Z$ independently of $y$. This leads to $S_1^{(1)}=2\sqrt{2}\cos\phi$ and $S_2^{(1)}=S_3^{(1)}=\sqrt{2}\left(\cos\phi+\sin\phi\right)$.

For $\lambda=2$, no party performs a unitary ($U=\openone$). Choose  $A_x=\cos\hat{\phi}\sigma_X+(-1)^x\sin\hat{\phi}\sigma_Z$, $B_0^{1}=\openone$ and $B_1^{1}=\sigma_Z$. $B_2$ and $B_3$ perform identical measurements $B_0^{2}=B_0^{3}=\sigma_X$ and $B_1^{2}=B_1^{3}=\sigma_Z$. This leads to $S_1^{(2)}=2\sin\hat{\phi}$, $S_2^{(2)}=\cos\hat{\phi}+2\sin\hat{\phi}$ and $S_3^{(2)}=\frac{\cos\hat{\phi}}{2}+\sin\hat{\phi}$. 

For $\lambda=3$, again all unitaries vanish $(U=\openone)$. Choose  $A_x=\cos\tilde{\phi}\sigma_X+(-1)^x\sin\tilde{\phi}\sigma_Z$. Let $B_1$ and $B_2$ measure $B_0^{1}=B_0^{2}=\openone$ and $B_1^{1}=B_1^{2}=\sigma_Z$, and $B_3$ measure $B_0^{3}=\sigma_X$ and $B_1^{3}=\sigma_Z$. This leads to $S_1^{(3)}=2\sin\tilde{\phi}$, $S_2^{(3)}=2\sin \tilde{\phi}$ and $S_3^{(3)}=\frac{1}{2}\left(\cos\tilde{\phi}+4\sin\tilde{\phi}\right)$. 

Now we obtain the final three CHSH parameters by mixing the three strategies. We have, for $k=1,2,3$, 
\begin{equation}
S_k=q_1 S_k^{(1)}+q_2 S_k^{(2)}+q_3 S_k^{(3)},
\end{equation}
One can find many choices of angles $(\phi,\hat{\phi},\tilde{\phi})$ and probabilities $q_\lambda=p(\lambda)$ such that $S_k>2$ for all three $k$. For example, insisting that $S_1=S_2=S_3\equiv S$ fixes $\{q_\lambda\}$ and one obtains $S=S(\phi,\hat{\phi},\tilde{\phi})$. Choosing $\left(\phi,\hat{\phi},\tilde{\phi}\right)=\left(\frac{31\pi}{132},\frac{88\pi}{245},\frac{16\pi}{33}\right)$, one obtains the triple violation $S\approx 2.00227$.

\end{document}